\documentstyle[twocolumn,eqsecnum,aps,amssymb]{revtex}

\newcommand{\be}{\begin{equation}}
\newcommand{\ee}{\end{equation}}
\newcommand{\bea}{\begin{eqnarray}}
\newcommand{\eea}{\end{eqnarray}}
\newcommand{\Ga}{\Gamma}
\newcommand{\half}{{\textstyle {1 \over 2}}}

\begin{document}
\draft
\preprint{UCL-IPT-99-02}
\title{Symmetric Boundary Conditions in Boundary Critical Phenomena}
\author{Philippe Ruelle\thanks{Chercheur qualifi\'e FNRS}}
\address{Universit\'e Catholique de Louvain\\
Institut de Physique Th\'eorique\\
B--1348 \hskip 0.5truecm Louvain-la-Neuve, Belgium
}
\date{\today}
\maketitle
\widetext
\begin{abstract}
Conformally invariant boundary conditions for minimal models on a cylinder are
classified by pairs of Lie algebras $(A,G)$ of ADE type. For each model,
we consider the action of its (discrete) symmetry group on the boundary
conditions. We find that the invariant ones correspond to the nodes in the
product graph $A \times G$ that are fixed by some automorphism. We 
proceed to determine the charges of the fields in the various Hilbert spaces,
but, in a general minimal model, many consistent solutions occur. In the unitary
models $(A,A)$, we show that there is a unique solution with the property that the
ground state in each sector of boundary conditions is invariant under the symmetry
group. In contrast, a solution with this property does not exist in the unitary
models of the series $(A,D)$ and $(A,E_6)$. A possible interpretation of
this fact is that a certain (large) number of invariant boundary conditions have
unphysical (negative) classical boundary Boltzmann weights. We give a tentative
characterization of the problematic boundary conditions. 
\end{abstract}

%\pacs{PACS numbers:\ 11.25.Hf,\ 05.50.+q,\ 75.10.Hk}

\narrowtext

%%%%%%%%%%%%%%%%%%%%%%%%%%%%%%%%%%%%%%%%%%%%%%%%%%%%%%%%%%%%%%%%%%%%%%%%%%%%%%

\section{Introduction}
\label{sec:intro}

It has been an extremely fruitful idea to study a conformal field theory by
putting it on various surfaces, with or without boundaries. Apart from the
sphere, that has been considered first, prime examples of non--trivial
geometries include the torus \cite{cardy1} and the cylinder
\cite{cardy2,cardy3}. They serve to probe different facets of a given conformal
theory. However the data specific of these surfaces are inextricably
related to each other, and this fact provides very stringent constraints on the
theory itself, allowing for example to determine its field content.

For minimal conformal theories, the problem on the torus for single--valued
fields has been resolved in \cite{ciz}: consistent models have a periodic
partition function that can be associated in a unique way with a pair
$(A,G)$ of simple Lie algebras of ADE type. 

The solution of the analogous problem for the cylinder is much more recent,
even if early calculations in either specific models or with specific boundary
conditions have been carried out in \cite{cardy2,cardy3,sb}. The recent
discovery in \cite{aos} of a new conformally invariant boundary condition in
the 3--state Potts model triggered a renewal of interest in the problem. For
minimal models, its solution was given in \cite{bpz,bppz}, and shown to be
encoded in the same Dynkin graphs that specify the torus partition function. 

When a model has a symmetry, necessarily discrete in this context, fields can be
multiple--valued on the torus, so that non--periodic sectors exist. Furthermore,
the fields transform under the symmetry group, and, upon diagonalization, can be
assigned charges.  All this information is encoded in frustrated partition\break

%\goodbreak

$$\mbox{}$$
\vspace{2.3cm}
\mbox{}

\noindent
functions, covariant under the modular group of the
torus, a fact that can be used to, first, detect the presence of a symmetry,
and then to compute the various partition functions \cite{zuber,rv}.

In this article, we address the question of the action of 
the symmetry group on the cylinder partition functions for the minimal
models. We show how the symmetry group acts on the boundary conditions, and
identify the invariant (or symmetric) ones. We then study the charge assignments
of the fields that occur in the presence of those boundary conditions. 

Section II is a reminder about the minimal conformal models on a torus and on a
cylinder. In Section III, we discuss the action of the symmetry group on the
conformally invariant boundary conditions, which is then used in Section IV to
compute frustrated partition functions on a cylinder, or equivalently the charge
assignment of the boundary fields. Section V contains explicit formulae and
computational details of a particular assignment. Its unicity (in fact non--unicity)
is examined in Section VI, from which we conclude that, in general, a large number
of distinct charge assignments are consistent. We also derive selection rules
for the boundary fusion coefficients. We finish, in Section VII, with an analysis
of the unitary models for which we propose an unambiguous charge assignment.

Section VII contains the most interesting corollary of the previous sections. An
analysis based on the expected consequences of the Perron--Frobenius theorem fixes
a unique charge assignment in the unitary $(A,A)$ models, which we conjecture to
be the correct one. This is in sharp contrast with the models of the $(A,D)$ and
$(A,E_6)$ series. For those, there is no consistent charge assignment that is
compatible with the Perron--Frobenius theorem, the reason being that there
is no way to ensure an invariant ground state in all sectors. Motivated by the
results obtained for the Potts model \cite{aos}, we will interpret this phenomenon
as the non--existence of positive classical Boltzmann weights for some invariant
boundary conditions. A simple characterization of them suggests itself in terms of
their Dynkin graph labels.

%%%%%%%%%%%%%%%%%%%%%%%%%%%%%%%%%%%%%%%%%%%%%%%%%%%%%%%%%%%%%%%%%%%%%%%%%%%%%%

\section{Minimal models}
\label{sec:rem}

Minimal models are classified by a pair $(A,G)$ of simply--laced
simple Lie algebras with coprime Coxeter numbers, $p$ and $q$. One may assume
that $p$ is odd. Their periodic partition function on a torus of modulus
$\tau$ is a sesquilinear form in the Virasoro characters
\be
Z(A,G) = \sum_{i,j} \; M_{ij}\,\chi^*_i(\tau)\,\chi^{}_j(\tau), \qquad M_{ij}
\in {\Bbb N},
\ee
where $i,j$ are labels for Virasoro highest weight representations. They lie in
the Kac table $\{(r,s)\;:\;1 \leq r \leq p-1,\, 1 \leq s \leq q-1\}$, in which
$(r,s)$ and $(p-r,q-s)$ must be identified. The connection with the Lie
algebras is best brought out by writing the diagonal elements $M_{ii}$ as
\cite{ciz}
\be
Z(A,G) = \half \sum_{r \in \text{Exp}\,A \atop s \in \text{Exp}\,G}\;
|\chi_{r,s}|^2 + \text{off-diagonal},
\label{zpp}
\ee
where $r$ and $s$ run over the exponents of $A$ and $G$. The full expressions
of the partition functions are given in \cite{ciz}. 

The question of the symmetry group has been first addressed in \cite{z}, and
solved in {rv} for the unitary models $|p-q|=1$. The analysis can however be easily
extended to the non--unitary minimal models, with the following result. With the
exception of the models $(A_{p-1},A_{q-1})$ with $p$ {\it and} $q$ odd, which have
no symmetry at all, the other models $(A,G)$ have a finite symmetry group $\Ga$,
which is the group of automorphisms of the Dynkin graph of $G$, that is,
$\Ga(G)=Z_2$ except
$\Ga(D_4)=S_3$ and $\Ga(E_7,E_8)=\{e\}$. 

When a model has a symmetry group, the fields may have a non--trivial monodromy
along the two periods of the torus, transforming as $\phi(z+1) = {}^{g}\phi(z)$
and $\phi(z+\tau) = {}^{g'}\!\phi(z)$ for two commuting\footnote{That forces us to 
focus on Abelian subgroups of $\Gamma$. Thus in this paper we consider
$Z_2$ and $Z_3$ (sub)groups only.} elements $g,g' \in \Ga$.
In the Hamiltonian formalism, this amounts to give a Hilbert space
${\cal H}_g$ of states with a $g$--monodromy along the first period, which are
then acted on by $g'$ when transported along the second period. The latter action
can be diagonalized, $\,^{g'}|\phi\rangle = e^{2i\pi Q/N} |\phi\rangle$, defining
the charge $Q$ of the field $\phi$ under the action of $g'$, an element of order
$N$. 

The field content of ${\cal H}_g$ as well as their charges can be read off from
the frustrated partition functions $Z_{g,g'}(A,G)$. These are still
sesquilinear forms but with coefficients in ${\Bbb Z}(e^{2i\pi/|\Ga|})$:
\be
Z_{g,g'} = {\rm Tr}_{{\cal H}_g} \;\Big[q^{L_0-c/24}\,\bar q^{\bar
L_0-c/24}\,g'\Big].
\ee
Because a modular transformation mixes the two periods, it must be accompanied
by a corresponding change of monodromies so that the net effect
vanish (for a fixed pair $(A,G)$):
\be
Z_{g,g'}(\tau) = Z_{g^ag'^c,g^bg'^d}({\textstyle{a\tau +b \over c\tau +d}}).
\label{tor}
\ee
All such functions are given explicitly in \cite{rv} (with a straightforward
extension to the non--unitary case).  The function (\ref{zpp}) corresponds to
$g=g'=e$.

On a cylinder, say of length $L$ and perimeter $T$, only one Virasoro algebra
remains, so that the partition function is linear rather than sesquilinear in
the characters \cite{cardy2}. Conformally invariant boundary conditions
$\alpha,\beta$ must be prescribed on the two boundaries, and a monodromy condition 
$g$ must be imposed along the periodic coordinate, $\phi(z+T)={}^g\phi(z)$. We
first consider a trivial monodromy, $g=e$.

If the time variable is defined to run along the periodic direction, the
partition function is the trace of the transfer matrix $e^{-TH_{\alpha,\beta}}$,
\be
Z^e_{\alpha,\beta}(\tau) = \sum_i \; n^i_{\alpha,\beta} \, \chi_i(\tau), \qquad 
\tau=iT/2L.
\label{form1}
\ee
The integer $n^i_{\alpha,\beta}$ gives the multiplicity of the primary field 
with Kac label $i$ in the Hilbert space ${\cal H}_{\alpha,\beta}$. 

Alternatively, one may view the time evolution as going from one boundary to the
other. In this case, the states on constant time surfaces belong to the bulk
periodic Hilbert space ${\cal H}_e$, and are propagated in time from one
boundary state $|\alpha\rangle$ to the other $|\beta\rangle$ (formally also in 
${\cal H}_e$). The partition function is then
\be
Z^e_{\alpha,\beta}(\tau) = \langle \beta| e^{-LH_e} |\alpha\rangle,
\label{preform2}
\ee
with $H_e$ the Hamiltonian corresponding to periodic bulk sector.

The boundary states are conformally invariant, satisfying $(L_n-\bar
L_{-n})|\alpha\rangle$ for all $n \in \Bbb Z$ \cite{cardy3}. The solutions to this
equation  are the Ishibashi states \cite{i}: every highest weight representation 
$[i \otimes \bar i]$ contains exactly one such state, which we denote by
$|i\rangle\rangle$, while the other representations $[i \otimes \bar j]$, for $i
\neq j$, do not contain any. In the present situation, the Ishibashi states
must be taken from the space ${\cal H}_e$, and hence are labelled by ${\cal
E}_e = \{i \;:\; [i \otimes \bar i] \in {\cal H}_e\}$. 

Expanding the boundary states in the basis of Ishibashi states, $|\alpha\rangle 
= \sum_i \; c^i_\alpha \, |i\rangle\rangle$, makes the partition function
(\ref{preform2}) take the form
\be
Z_{\alpha,\beta}(\tau) = \sum_{i \in {\cal E}_e} \; c^i_\alpha \, 
\bar c^{i}_\beta \, \chi_i({\textstyle{-1 \over \tau}}).
\label{form2}
\ee

The arguments of the characters in (\ref{form1}) and (\ref{form2}) are related
by the modular transformation $\tau \mapsto {-1 \over \tau}$, under which the
characters transform linearly through a unitary matrix $S$. Comparing the two
formulae then yields Cardy's equation \cite{cardy3}
\be
n^i_{\alpha,\beta} = \sum_{j \in {\cal E}_e} \; S_{i,j} \, c^j_\alpha \, 
\bar c^{j}_\beta.
\label{cyl}
\ee

The relations (\ref{cyl}) are overdetermined for the vectors $c^j$, and
provide a means to classify the boundary conditions $|\alpha\rangle$, to compute
the spectra of ${\cal H}_{\alpha,\beta}$, and in turn the surface scaling 
dimensions. Such calculations have been carried out in \cite{cardy2,sb,aos},
but the general answer appeared only very recently in \cite{bpz,bppz}. Let 1 be
the label corresponding to the vacuum representation, namely to
$(r,s)=(1,1)=(p-1,q-1)$.

In \cite{bppz}, it was observed that, upon setting $c^i_\alpha =
\psi^i_\alpha / \sqrt{S_{1,i}}$ for a set of complete and orthonornal vectors
$\psi^i$, Cardy's equation appears as an explicit diagonalization 
\be
n^i_{\alpha,\beta} = \sum_{j \in {\cal E}_e} \; \psi^j_\alpha \, {S_{i,j} \over 
S_{1,j}} \, \bar\psi_\beta^{j}.
\label{nie}
\ee
The matrices $n^i$ have eigenvalues ${S_{i,j}/S_{1,j}}$, and a
common eigenbasis is given by the vectors $\psi^j$. As a result, they satisfy
the fusion rules
\be
n^i \, n^j = \sum_k \; N_{ij}^k \, n^k.
\ee
Reversing the argument, the authors of \cite{bppz} conclude that an $\Bbb
N$--valued representation of the fusion algebra of dimension $|{\cal
E}_e|$ provides a solution to Cardy's equation with $|{\cal E}_e|$ different
boundary conditions. When $c^i_\alpha = \psi^i_\alpha / \sqrt{S_{1,i}}$ is 
an invertible matrix, this solution yields the maximal set of
conformally invariant boundary conditions. Note that the boundary states
$|\alpha\rangle$ are determined up to a phase, but the fact that the entries of 
$n^i$ are to be positive integers leaves only a global, unobservable, phase.

For minimal models, this was all made explicit in \cite{bpz}. For the model
$(A,G)$, it was shown that each node in the product Dynkin
diagram $A \times G$, quotiented by an appropriate $Z_2$ automorphism, 
defines a boundary condition and vice--versa. Indeed, from (\ref{zpp}), the
number of Ishibashi states in the periodic sector is $|{\cal E}_e| = \half
|\text{Exp} A \times \text{Exp} G|$, so that only half the nodes can define
distinct boundary conditions. We will use the variables $\alpha,\beta$
and $(a_i,b_i)$ as labels for the nodes of $A \times G$. The letters $A$ and $G$
will denote at the same time the Lie algebras, the Dynkin diagrams or the
corresponding adjacency matrices.

As a result of the quotient of the product graph, the matrices $n^i$,
for $i=(r,s)$, are given by \cite{bpz}
\bea
n^i_{(a_1,b_1),(a_2,b_2)} &=& (\hat N_r)_{a_1,a_2}\,(V_s)_{b_1,b_2} + 
(\hat N_r)_{a_1,a^*_2}\,(V_s)_{b_1,b^*_2} \nonumber \\
&=& n^i_{(a^*_1,b^*_1),(a_2,b_2)} = n^i_{(a_1,b_1),(a^*_2,b^*_2)}.
\label{ni}
\eea
In this formula, the $\hat N$'s and the $V$'s are the fused adjacency matrices
of $A$ and $G$ respectively. They are defined recursively by $X_m = X_2X_{m-1} -
X_{m-2}$, with $X_1={\bf 1}$ and $X_2=A$ if $X=\hat N$, and $X_2=G$ if $X=V$. 
Equivalently, 
\be
\hat N_r = U_{r-1}(A), \qquad V_s = U_{s-1}(G), 
\ee
where
$U_m(2\cos{x}) = \sin{(m+1)x}/\sin{x}$ is the $m$--th Tchebychev polynomial of
the second kind. The automorphism $(a,b) \mapsto (a^*,b^*)$ can be determined 
from the condition $n^{(r,s)}=n^{(p-r,q-s)}$ (necessary if the $n^i$ are to
satisfy the fusion algebra). It yields $a^*$ and $b^*$ to be given\footnote{The
automorphism $^*$ in $G$ thus  coincides with the charge conjugation in the
corresponding affine algebra $\hat G$.} by the non--trivial automorphism of $A$
and $G$, for $G \neq D_{\text{even}},E_7,E_8$, and $b^*=b$ for
$G=D_{\text{even}},E_7,E_8$. 

Viewing the tensor products $F^i(A,G) = \hat N_r \otimes V_s$ as the fused
adjacency matrices of $A \times G$, the above result may be summarized by
saying that $n^i$ is a folded fused adjacency matrix of $A \times G$
\be
n^i_{\alpha,\beta} = F^i_{\alpha,\beta}(A,G) + F^i_{\alpha,\beta^*}(A,G).
\label{nif}
\ee

The eigendata for the matrices $A$ and $G$ make sure that the matrices in
(\ref{ni}) satisfy the minimal model fusion algebra. For the $(A,A)$ models,
the $a_i$ (resp. $b_i$) labels run over the same set as $r$ (resp. $s$), and the
matrices $n^i$ are the fusion matrices $N^i$ themselves \cite{cardy3}. 

%%%%%%%%%%%%%%%%%%%%%%%%%%%%%%%%%%%%%%%%%%%%%%%%%%%%%%%%%%%%%%%%%%%%%%%%%%%%%%

\section{Symmetric boundary conditions}
\label{sec:sym}

We now proceed to the analysis of the cylinder partition functions when there
is a group of symmetry $\Ga$. From now on, we thus take $q$ even, and $G
\neq E_7,E_8$.

The boundary states are combinations of periodic Ishibashi states, on which
the action of $\Ga$ is known from the torus partition functions $Z_{e,g}$.
This induces an action on the boundary states which one can determine.
That action must be by permutations. 

For the minimal models, a boundary state corresponds to a pair of nodes of $A$
and $G$,
\be
|(a,b)\rangle = \sum_{i \in {\cal E}_e} \; {1 \over
\sqrt{S_{1,i}}} \,\psi^i(a,b)\, |i\rangle\rangle.
\label{exp}
\ee
where the $\psi^i$ form an eigenbasis for the concrete matrices in (\ref{ni}).

Let us denote by $\sigma$ the automorphisms of the Dynkin graph of $G$, so that
every $\sigma$ has fixed points. (The automorphism of the $A$ factor has a free
action, and is used to obtain a set of representatives under the $^*$ involution,
see (\ref{ni}).) Each $\sigma$ has a diagonalizable action on the eigenvectors
$\psi^i$.

The action of $g \in \Ga$ on a periodic Ishibashi state can be read off from
the diagonal terms in the frustrated partition function $Z_{e,g}(A,G)$
\cite{rv}. These can be compactly presented as follows. If $g$ has order $N$,
and if one writes the diagonal terms in $Z_{e,g}$ as 
\be
Z_{e,g} = \sum_{i \in {\cal E}_e} \; \zeta_{N}^{Q_{g}(i)} \: |\chi_i|^2 +
\ldots,
\ee
then, for a proper choice of the $\psi^i$, the phase is seen to be exactly
equal to the eigenvalue of $\psi^i$ under an order $N$ automorphism $\sigma$:
\be
\psi^i(a,\sigma(b)) = \zeta_{N}^{Q_{g}(i)} \, \psi^i(a,b).
\label{auto}
\ee
The $\sigma$ that is induced by $g$ through the previous formula is unambiguous
in the models $(A,G)$ if $G$ is not $D_4$: the only non--trivial $g$
induces the only non--trivial $\sigma$. When the $D_4$ algebra is involved,
exactly which $\sigma$ in $S_3$ arises from a set of charges $Q_g$ (univoquely
given by $Z_{e,g}$) depends on the eigenbasis we choose. In particular, a same
set of $Z_2$ charges can lead to the three different (but conjugate) order two
$\sigma$'s. 

It quickly follows from (\ref{exp}) and (\ref{auto}) that an order $N$ group element
$g$ acts on the boundary states as an order $N$ automorphism $\sigma$:
\be
|(a,b)\rangle \longrightarrow |^g(a,b)\rangle = |(a,\sigma(b))\rangle.
\ee
Therefore, for any subgroup $\gamma$ of $\Ga$, the $\gamma$--symmetric boundary
conditions correspond to the nodes of $A \times G$ that are fixed by
a group $\gamma$ of automorphisms of $G$. This set of nodes form a graph which
we call the fixed point graph and denote by $A \times G^\gamma$.

In particular, the boundary conditions that are invariant under a group
element $g$ correspond to the nodes in $A \times G^\sigma$, with
$G^\sigma$ the part of $G$ that is fixed by the automorphism $\sigma$ induced
by $g$. As before the pairs of nodes which are related by the $^*$ automorphism
define the same invariant boundary conditions. In the minimal models, the fixed
point diagrams that arise for the various choices of
$g$ are 
\bea
(A_{p-1},A_{q-1})\;:\quad && T_{(p-1)/2} \times A_1,\nonumber \\
(A_{p-1},D_{q/2+1})\;:\quad && T_{(p-1)/2} \times A_{q/2-1}, \quad
(g^2=e), \nonumber \\ 
(A_{p-1},D_4)\;:\quad && T_{(p-1)/2} \times A_1, \quad (g^3=e),\nonumber \\ 
(A_{p-1},E_6)\;:\quad && T_{(p-1)/2} \times A_2,
\label{fixdyn}
\eea
where $T_{(p-1)/2}$ denotes the tadpole diagram obtained by quotienting
$A_{p-1}$ by its automorphism $^*$. 

For instance, the fixed point graph of an element $g$ of order two in the
$(A_{p-1},D_{q/2+1})$ model is graphically given by

\hskip 1truecm
\begin{picture}(180,35)(0,-10)
\put(-4,0){\circle*{4}}
\put(-10,-10){\makebox(0,0)[b]{\small $a\!=\!1$}}
\put(-2,0){\line(1,0){15}}
\put(15,0){\circle*{4}}
\put(15,-10){\makebox(0,0)[b]{\small 2}}
\put(17,0){\line(1,0){10}}
\put(30,0){...}
\put(40,0){\line(1,0){10}}
\put(52,0){\circle*{4}}
\put(52,-10){\makebox(0,0)[b]{\small $p-1 \over 2$}}
\put(52,7){\circle{14}}

\put(83,0){\makebox(0,0)[c]{$\times$}}

\put(110,0){\circle*{4}}
\put(104,-10){\makebox(0,0)[b]{\small $b\!=\!1$}}
\put(112,0){\line(1,0){15}}
\put(129,0){\circle*{4}}
\put(129,-10){\makebox(0,0)[b]{\small 2}}
\put(131,0){\line(1,0){10}}
\put(144,0){...}
\put(154,0){\line(1,0){10}}
\put(166,0){\circle*{4}}
\put(166,-10){\makebox(0,0)[b]{\small ${q \over 2}-1$}}
\end{picture}
\vskip 0.4truecm

%%%%%%%%%%%%%%%%%%%%%%%%%%%%%%%%%%%%%%%%%%%%%%%%%%%%%%%%%%%%%%%%%%%%%%%%%%%%%%

\section{Cylinder partition functions}
\label{sec:cpf}

The consequences of a symmetry can now be pursued at the level of the partition
functions. Let us suppose that $\alpha$ and $\beta$ are two boundary conditions 
that are invariant under a group element $g$, of order $N$. 

It implies that the transfer matrix $e^{-H_{\alpha,\beta}}$ and $g$ commute, and 
can be diagonalized in the same basis. The effect, on the cylinder partition
function, of the insertion of $g$ on a line connecting the two boundaries is to
affect each Virasoro tower with a $N$--th root of unity, so that the first form
(\ref{form1}) becomes
\be
Z^g_{\alpha,\beta}(\tau) = \sum_i \; n^{(g)\,i}_{\alpha,\beta} \, \chi_i(\tau).
\ee
This shows that $n^{(g)\,i}$ must be related in the following way to the
restriction of $n^i$ to the $g$--symmetric boundary conditions: an entry of $n^i$
equal to $n$ becomes in $n^{(g)\,i}$ a sum of $n$ $N$--th roots of unity. 

In the second form, the boundary state $|\alpha\rangle$ is propagated to 
$|\beta\rangle$  by the Hamiltonian that acts on the bulk sector twisted by $g$,
so that
\be
Z^g_{\alpha,\beta}(\tau) = \langle \beta| e^{-LH_g} |\alpha\rangle.
\ee
This formula makes it clear that the states $|\alpha\rangle$ and
$|\beta\rangle$ should have a projection in the twisted Hilbert space ${\cal
H}_g$, and being conformally invariant, must have expansions in Ishibashi
states of the bulk $g$--sector, themselves labelled by ${\cal E}_g = \{i \;:\;
[i \otimes \bar i] \in {\cal H}_g\}$. Setting $|\alpha\rangle = \sum_i \;
c^{(g)\,i}_\alpha \, |i\rangle\rangle_g$, one obtains a Cardy equation 
\be
n^{(g)\,i}_{\alpha,\beta} = \sum_{j \in {\cal E}_g} \; S_{i,j} \,
c^{(g)\,j}_\alpha \, \bar c^{(g)\,j}_\beta,
\label{cylt}
\ee
for all boundary conditions which are $g$--symmetric. 

By inspecting the torus partition functions $Z_{g,e}(A,G)$ \cite{rv} (see also
the next section), one readily sees that the matrices $c^{(g)\,i}_\alpha$ are
square, namely 
\be
|{\cal E}_g| = \half \, |A \times G^\sigma| = |T \times G^\sigma|,
\ee
where the factor $\half$ accounts for the identification under $^*$.
Let us also note that, since the $g$--Ishibashi states form a basis for boundary
states that are invariant under $g$, they should themselves be all neutral for
consistency. This is again easily checked from $Z_{g,g}$. 

The rest of this article is devoted to a discussion of the solutions to the
Cardy equation (\ref{cylt}). We will suggest that the proper
physical solution is a natural generalization to $g \neq e$ of the two
formulae (\ref{nie}) and (\ref{nif}) for $n^i$. 

Our first statement is that a particular solution, compatible with $n^i \equiv
n^{(e)\,i}$, is provided, modulo a sign $\delta_i$, by the folded fused
adjacency matrices of the graph $A \times G^\sigma$:
\be
\tilde n^{(g)\,i}_{\alpha,\beta} = \delta_i \, \big[F^i_{\alpha,\beta}
(A,G^\sigma) + F^i_{\alpha,\beta^*}(A,G^\sigma)\big], \quad \delta_i=\pm 1.
\label{fff}
\ee
Here $\alpha=(a_1,b_1)$ and $\beta=(a_2,b_2)$ are pairs of nodes in $A \times
G^\sigma$ (with the usual identification under $^*$), and the automorphism $^*$
is the same as before. 

When $g,\sigma \neq e$, this formula can be simplified because every $b_2$ in
$G^\sigma$ is a fixed point of $^*$. Indeed since $\beta$ is a node of $A \times
G^\sigma$, $b_2$ is a fixed point of $\sigma$. But $\sigma$ and $^*$ coincide,
except for $G=D_{\text{even}}$ for which $^*$ is trivial. Thus the folding
by $^*$  acts on $a_2$ only, resulting in an effective folding of the $A$ factor
onto a $T$ graph (hence the graphs (\ref{fixdyn})). One also checks that the
folded fused adjacency matrices of $A_{p-1}$ are the fused adjacency matrices of
$T_{(p-1)/2}$. Thus the matrices in (\ref{fff}) are simply proportional to the
fused adjacency matrices of the fixed point diagram 
\be
\tilde n^{(g)\,i}_{\alpha,\beta} = \delta_i \, F^i_{\alpha,\beta}(T,G^\sigma)
= \delta_i \, U_{r-1}(T)_{a_1,a_2} U_{s-1}(G^\sigma)_{b_1,b_2}.
\label{nigf}
\ee

The matrices $F^i(T,G^\sigma)$ fall short of satisfying the minimal fusion
algebra, but the factors $\delta_i$ can be adjusted so that the $\tilde n^{(g)\,i}$ do
satisfy it.

The fusion algebra of the minimal model ${\cal M}(p,q)$ is polynomially generated
by two generators $X$ and $Y$, which one can associate with the representatives of
$N^{(2,1)}$ and $N^{(1,2)}$ \cite{dms}. The other elements of the algebra are
explicitly given by Tchebychev polynomials
\be
N^i = U_{r-1}(X) \, U_{s-1}(Y),
\label{pform}
\ee
and the generators must satisfy three relations:
\be
U_{p-1}(X) = U_{q-1}(Y) = U_{p-2}(X) - U_{q-2}(Y) = 0.
\label{rel}
\ee

The matrices $F^i(T,G^\sigma)$ have the proper form (\ref{pform}), and 
$T_{(p-1)/2}$ and $G^\sigma$ do indeed satisfy the first two
relations in (\ref{rel}). This is most easily seen by verifying that all
eigenvalues satisfy the relevant equation. For instance, the eigenvalues
$\lambda_m$ of $T_{(p-1)/2}$ are in
\be
\text{spec}(T_{{p-1\over 2}}) = \{2\cos{\textstyle{\pi m \over
p}}\;:\; 1 \leq m\,\text{odd} \leq p-1\},
\label{spectrum}
\ee
and clearly satisfy $U_{p-1}(\lambda_m)=0$. 

In the same way, one computes that
\be
U_{p-2}(T_{{p-1\over 2}}) = {\bf 1}.
\ee
The corresponding calculation for $G^\sigma$ yields\footnote{The adjacency matrix
of $A_1$ is the number zero, so that its fused adjacency matrices are $U_{s-1}(0)
= (-1)^{(s-1)/2}$ for $s$ odd, and 0 for $s$ even.}, in the same four cases as in
(\ref{fixdyn}),
\bea
G^\sigma=A_1\;: && \qquad U_{q-2}(G^\sigma) = (-1)^{{q \over 2}+1}\,{\bf 1},
\nonumber\\ 
G^\sigma=A_{{q \over 2}-1}\;: && \qquad U_{q-2}(G^\sigma) = -{\bf 1}, \nonumber\\
G^\sigma=A_1\;: && \qquad U_{q-2}(G^\sigma) = {\bf 1},\nonumber\\
G^\sigma=A_2\;: && \qquad U_{10}(G^\sigma) = -{\bf 1},
\label{ug}
\eea
where the last line refers to the models $(A_{p-1},E_6)$ for which $q=12$.
Thus, except when $G^\sigma=A_1$ and when $q=2 \bmod 4$, the last condition in
(\ref{rel}) is not fulfilled. 

Owing to the parity properties of the Tchebychev polynomials, $U_m(-x) = (-1)^m
U_m(x)$, one easily sees that $X=(-1)^{{q \over 2}+1}\,T_{(p-1)/2}$ in the first
and third cases of (\ref{ug}), and $X=-T_{(p-1)/2}$ in the second and fourth ones,
together with $Y=G^\sigma$, do satisfy all three conditions and therefore
generate the correct algebra. 

Correspondingly, one finds that the matrices $\tilde n^{(g)\,i} = F^i(X,Y) = 
\delta_i F^i(T,G^\sigma)$ with the following signs,
\bea
(A_{p-1},A_{q-1}) \;: && \quad \delta_i = (-1)^{(r+1)({q \over
2}+1)},\nonumber\\
(A_{p-1},D_{{q \over 2}+1})\;: && \quad \delta_i = (-1)^{r+1}, \qquad
(g^2=e),\nonumber\\ (A_{p-1},D_4)\;: && \quad \delta_i = 1, \qquad (g^3=e),
\nonumber\\ (A_{p-1},E_6)\;: && \quad \delta_i = (-1)^{r+1}.
\label{deltas}
\eea
obey the minimal fusion algebra. Because of the signs $\delta_i$ but also because
the matrices $F^i(T,G^\sigma)$ are not positive for $\sigma \neq e$ (they are
however of constant sign), the $\tilde n^{(g)\,i}$ provide $\Bbb
Z$--representations\footnote{In case of a $Z_3$ symmetry group, one might expect
${\Bbb Z}(e^{2i\pi/3})$--valued representations. This is however excluded by the
symmetry $Z^g_{\alpha,\beta} = Z^g_{\beta,\alpha}$ (time reversal invariance),
which implies the reality of $n^{(g)\,i}_{\alpha,\beta}$.} of the minimal fusion
algebra. 

It remains to prove our earlier assertion that the so--defined $\tilde n^{(g)\,i}$ 
are solutions to Cardy's equation (\ref{cylt}). 

Since they satisfy the fusion algebra, the $\tilde n^{(g)\,i}$ must have 
eigenvalues given by ratios ${S_{i,j} \over S_{1,j}}$ of $S$ matrix elements. It
is not difficult to see, by looking first at the partition functions $Z_{g,e}$ to
get ${\cal E}_g$ and then by computing the ratios explicitly, that the eigenvalues
of $\tilde n^{(g)\,i}$ are precisely the above ratios for $j \in  {\cal E}_g$ (see
next section). Thus the following diagonalization formulae hold
\be
\tilde n^{(g)\,i}_{\alpha,\beta} = \sum_{j \in {\cal E}_g} \; \psi^{(g)\,j}_\alpha \,
{S_{i,j} \over S_{1,j}} \, \bar\psi_\beta^{(g)\,j},
\label{nig}
\ee
where the vectors $\psi^{(g)\,j}$ form a common orthonormal eigenbasis
(also common to all fused adjacency matrices $F^i(T,G^\sigma)$ of the fixed point
diagram). This yields the value of the coefficients in (\ref{cylt})
\be
c^{(g)\,j}_\alpha = {1 \over \sqrt{S_{1,j}}}\,\psi^{(g)\,j}_\alpha.
\label{coeff}
\ee

To complete the proof, it is enough to show that they are compatible with the
$n^i$, in the sense that has been explained in Section \ref{sec:sym}: an entry in
$n^i$ equal to $n$ goes over, in $\tilde n^{(g)\,i}$, to a sum of $n$ roots of 
unity, and moreover $\tilde n^{(g)\,1} = {\bf 1}$. One may verify that this is 
indeed the case. We omit the proof here since, to a large extent, it is given in
the next section.

The formulae (\ref{fff}) and (\ref{nig}) bear much resemblance
with the corresponding ones for $n^i$, of which they constitute a natural
extension. Like the $n^i$, the matrices $\tilde n^{(g)\,i}$ have a graph theoretic
description derived from that of $n^i$ through the action of $g$, they satisfy 
the minimal fusion algebra, and their eigenvalues are exactly labelled by the  
set ${\cal E}_g$ which specifies the diagonal terms of the twisted partition
functions $Z_{g,e}$. In a sense, this set ${\cal E}_g$ can also be viewed as the 
set of exponents of the fixed point graph that serves to define $\tilde n^{(g)\,i}$.

%%%%%%%%%%%%%%%%%%%%%%%%%%%%%%%%%%%%%%%%%%%%%%%%%%%%%%%%%%%%%%%%%%%%%%%%%%%%%%

\section{More explicit formulae}
\label{sec:expl}

We give here the computational details and the proofs that were missing in the
previous section. 

We begin by recalling the formula giving the $S$ matrix elements in
the minimal model ${\cal M}(p,q)$, for $i=(r,s)$ and $j=(r',s')$,
\be
S_{i,j} = \sqrt{8 \over pq} \, (-1)^{rs'+r's+1} \, \sin{\pi qrr' \over p} \,
\sin{\pi pss' \over q}.
\ee
We examine in turn each of the three infinite series.

\subsection{The series (A,A)}

The models $(A_{p-1},A_{q-1})$, $p$ odd and $q$ even, have the symmetry group
$Z_2$. The invariant boundary conditions $\alpha=(a,b)$ are controlled by the
tadpole graph $T_{(p-1)/2} \times A_1$, i.e. $a$ runs from 1 to $(p-1)/2$ and
$b=q/2$. 

The frustrated partition functions are \cite{rv},
\be
Z_{g,e}(A,A) = \half \sum_{r,s} \chi_{r,s}^* \, \chi^{}_{r,q-s} = \sum_{1 \leq
r\;\text{odd} \leq p-1 \atop 1 \leq s  \leq q-1} \; \chi_{r,s}^* \,
\chi^{}_{r,q-s},
\ee
from which it follows that the twisted Ishibashi states $|j\rangle\rangle_g$ can
be labelled by 
\be
{\cal E}_g(A,A) = \{j=(m,{\textstyle {q \over 2}}) \;:\; 1 \leq m \;{\rm odd} 
\leq p-1\}.
\ee
(Which representative $(r,s)$ or $(p-r,q-s)$ we take does not matter, since the
$S$ matrix elements are the same.)

For these values of $j$, an easy calculation yields
\be
{S_{i,j} \over S_{1,j}} = (-1)^{(r+1)({q \over
2}+1)}\,U_{r-1}(-2\cos{\textstyle {\pi qm \over p}})\,U_{s-1}(0).
\label{ratios1}
\ee
Since $q$ is even, the numbers which appear as arguments of $U_{r-1}$ coincide
with the set (\ref{spectrum}) of eigenvalues of the incidence matrix
$T_{(p-1)/2}$. A simple comparison with the matrices $\tilde n^{(g)\,i}$, as
computed from (\ref{nigf}) and (\ref{deltas}), 
\be
\tilde n^{(g)\,i} = (-1)^{(r+1)({q \over 2}+1)} \, U_{r-1}(T_{p-1 \over
2}) \, U_{s-1}(0).
\ee
shows that the eigenvalues of $\tilde n^{(g)\,i}$ are indeed the numbers in
(\ref{ratios1}) for $j \in {\cal E}_g$.

As mentioned before, the matrices $n^i$ are the fusion matrices $N^i$
themselves \cite{cardy3}, equal, from (\ref{ni}), to
\be 
n^i_{(a_1,{q \over 2}),(a_2,{q \over 2})} =  N^i_{(a_1,{q \over 2}),(a_2,{q \over
2})} =  U_{r-1}(T_{{p-1\over 2}})_{a_1,a_2},
\ee
for all odd $s$, and identically equal to zero for $s$ even. This then leads to 
\be
\tilde n^{(g)\,i}_{(a_1,{q \over 2}),(a_2,{q \over 2})} = (-1)^{(r+1)({q \over
2}+1)+{s-1 \over 2}} \, N^i_{(a_1,{q \over 2}),(a_2,{q \over 2})}.
\label{aa}
\ee
This equation shows clearly that $\tilde n^{(g)\,i}$ is compatible with $n^i$ in
the sense explained before.

\subsection{The series (A,D)}

All models $(A_{p-1},D_{q/2+1})$, with two coprime integers $p,q$ and $p$ odd as
before, have also a $Z_2$ symmetry. The non--trivial group element $g$ induces the
automorphism $\sigma$ of $D_{q/2+1}$ which exchanges the last two nodes.
Therefore the symmetric boundary states correspond to the nodes $(a,b)$ of the
fixed point diagram $T_{(p-1)/2} \times A_{q/2-1}$, pictured in Section
\ref{sec:sym}, so that $a$ is between 1 and $(p-1)/2$, and $b$ is between 1
and $q/2-1$.

The eigenvalues of $T_{(p-1)/2}$ have been recalled earlier, while those of
$A_{q/2-1}$ are well known:
\bea
&& \text{spec}(T_{{p-1\over 2}}) = \{2\cos{\textstyle{\pi m \over p}}\;:\; 1
\leq m\,\text{odd} \leq p-1\}, \label{spect} \\
&& \text{spec}(A_{{q \over 2}-1}) = \{2\cos{\textstyle{\pi m' \over q}}\;:\; 1
\leq m'\,\text{even} \leq q-1\}.
\label{speca}
\eea

The frustrated (antiperiodic) partition function on the torus is (the
double sums run over $[1,p-1]\times[1,q-1]$) \cite{rv}
\be
Z_{g,e}(A,D) = \sum_{r \, {\rm odd} \atop s \,{\rm even}} |\chi^{}_{r,s}|^2 + 
\sum_{r \, {\rm odd} \atop s=1+{q \over 2} \bmod 2} \chi_{r,s}^*\chi^{}_{r,q-s}.
\ee
Thus the Kac labels of the $g$--Ishibashi states $|j\rangle\rangle_g$ can be
chosen in the set 
\bea
{\cal E}_g(A,D) &=& \{j=(m,m') \;:\; 1 \leq m \;{\rm odd} \leq p-1, \nonumber\\
&& \hskip 0.6cm 1 \leq m' \;{\rm even} \leq q-1\}.
\eea

{}From this, one computes
\be
{S_{i,j} \over S_{1,j}} = (-1)^{r+1} \, U_{r-1}(-2\cos{\textstyle{\pi qm
\over p}}) \,  U_{s-1}(-2\cos{\textstyle{\pi pm' \over q}}),
\label{rat}
\ee
which coincide, in view of (\ref{spect}) and (\ref{speca}), with the eigenvalues
of 
\be
\tilde n^{(g)\,i}_{\alpha,\beta} = (-1)^{r+1} \:
U_{r-1}(T_{{p-1\over 2}})_{a_1,a_2} \, U_{s-1}(A_{{q \over 2}-1})_{b_1,b_2}.
\label{ad}
\ee

The numbers in the set $\{2\cos{\textstyle{\pi pm' \over
q}}\}$ come by pairs of opposite sign, so that the set of ratios (\ref{rat}), for
fixed $i$, is the same whether or not there is a minus sign in the argument of
$U_{s-1}$. Each individual ratio however differs by a factor $(-1)^{s+1}$, which
then leads to an alternative solution $(-1)^{s+1} \tilde n^{(g)\,i}$.

Finally the compatibility of $\tilde n^{(g)\,i}$ with the original matrices $n^i$ 
can be established. In the sector of invariant boundary conditions, the latter
read 
\be
n^i_{\alpha,\beta} = U_{r-1}(T_{{p-1\over 2}})_{a_1,a_2} \, U_{s-1}(D_{{q \over
2}+1})_{b_1,b_2},
\ee
where $b_1,b_2$ are restricted to lie in $[1,q/2-1]$. One may
simply notice the following modular identity (same values of the indices)
\be
U_{s-1}(D_{{q \over 2}+1}) = U_{s-1}(A_{{q \over 2}-1}) \bmod 2.
\ee
It has the immediate consequence that 
\be
\tilde n^{(g)\,i}_{\alpha,\beta} = n^i_{\alpha,\beta} \bmod 2,
\ee
which shows the required compatibility.

One may note that all the entries of $\tilde n^{(g)\,i}$ are in $\{0,+1,-1\}$,
and that those of $n^i$ are in $\{0,1,2\}$, which implies that all doubled
primary fields have opposite $Z_2$ charges within each pair. 

When $q=6$, i.e. for the $(A_{p-1},D_4)$ models, $Z_3$ invariant boundary
conditions can be investigated. They are labelled by nodes $(a,2)$ with $a$ in
$T_{(p-1)/2}$. 

The $Z_3$ frustrated partition functions on the torus are \cite{rv}
\be
Z_{g,e}(A,D_4) = \sum_{r\,{\rm odd}} |\chi^{}_{r,3}|^2 + \sum_{r\,{\rm odd}}
\chi^*_{r,3}[\chi^{}_{r,1}+\chi^{}_{r,5}]+{\rm c.c.},
\ee
so that the Ishibashi states in the $Z_3$--twisted sector have labels $j=(m,3)$
for $m$ odd between 1 and $p-1$. 

The matrices $\tilde n^{(g)\,i}$ in (\ref{nigf}) can be compared with the
restriction of $n^i$ to the sector of invariant boundary conditions, given by
$U_{r-1}(T_{(p-1)/2})_{a_1,a_2} U_{s-1}(D_4)_{2,2}$. All matrices are identically
zero for $s$ even, while for $s$ odd:
\bea
&& n^i = U_{r-1}(T_{{p-1\over 2}}) = \tilde n^{(g)\,i},
\quad \text{for}\;s=1,5,\nonumber\\
&& n^i = 2U_{r-1}(T_{{p-1\over 2}}), \quad 
\tilde n^{(g)\,i} = -U_{r-1}(T_{{p-1\over 2}}), \quad \text{for}\;s=3. \nonumber\\
\label{z3}
\eea
As in the $Z_2$ case, the second line shows that the doubled fields have
opposite and non--zero $Z_3$ charge (if $\omega \neq 1$ is a third root of unity,
$\omega +\omega^2=-1$).

\subsection{The series (A,E$_{\bf 6}$)}

The models $(A_{p-1},E_6)$ are similar to the $(A,D)$ models. In particular the
formula for the matrices $\tilde n^{(g)\,i}$ is the same as for the $(A,D)$ models
(with $A_{q/2-1}$ replaced by $A_2$). 

A unique feature of the models based on $E_6$ however is that some of the fields
occur tripled in some boundary conditions (in addition to some others being
doubled). One finds that these are the fields $(r,s)$ with $s=5$ and 7, in the
boundary conditions corresponding to the nodes $(a,3)$, for $a$ in
$T_{(p-1)/2}$ (with $b=3$ the intersection of the three branches of $E_6$). This
follows from the fused adjacency matrices $U_4(E_6)$ and $U_6(E_6)$, which, when
restricted to the nodes $b=3,6$ corresponding to the symmetric boundary conditions,
read
\be
U_4(E_6) = U_6(E_6) = \pmatrix{3 & 0 \cr 0 & 1}.
\ee

%%%%%%%%%%%%%%%%%%%%%%%%%%%%%%%%%%%%%%%%%%%%%%%%%%%%%%%%%%%%%%%%%%%%%%%%%%%%%%

\section{Unicity}

The boundary conditions that are invariant under a group element $g$
correspond to boundary states which have expansions in $g$--Ishibashi
states \footnote{The full expansion of $|\alpha\rangle$ involves Ishibashi states
from the $g$--twisted bulk sectors for all $g$ which leave $\alpha$ invariant.}
\be
|\alpha\rangle = \sum_{i \in {\cal E}_g} \; c^{(g)\,i}_\alpha \,
|i\rangle\rangle_g.
\ee
The coefficients in (\ref{coeff}) provide a specific solution $\tilde n^{(g)\,i}$ 
to Cardy's equation (\ref{cylt}). As for the $n^i$, one may raise the question of
the unicity of this solution. 

For every $g$, the symmetric boundary conditions exhaust the $g$--Ishibashi
states. It means that every other symmetric boundary state must be a linear
combination of those we already have, and so must be one of them. However, since
the boundary states $|\alpha \rangle$ enter Cardy's formula through scalar
products, it is the boundary rays more than the boundary states which matter.
Thus the basic question is whether one keeps a sensible solution if one
multiplies the boundary states by phases.

Clearly if the symmetric boundary states are multiplied by phases,
$|\alpha\rangle \rightarrow \varphi_\alpha |\alpha\rangle$, the matrices change
according to $\tilde n^{(g)\,i}_{\alpha,\beta} \rightarrow \varphi_\alpha
\varphi^*_\beta \tilde n^{(g)\,i}_{\alpha,\beta}$, which satisfy the minimal 
fusion algebra for any choice of phases. 

Whereas for $g=e$, the positivity of $n^{(e)\,i}=n^i$ forces all the phases to
be equal, this is no longer the case when $g \neq e$. Since the matrices
$n^{(g)\,i}$ are $\Bbb Z$--valued, the only condition one has is that the
phases must be equal up to signs, $\varphi_\alpha=\epsilon_\alpha \, \varphi$. 

For a $Z_2$ symmetry (or subgroup), the new matrices $\epsilon_\alpha
\epsilon_\beta \tilde n^{(g)\,i}_{\alpha,\beta}$ are also solutions of the Cardy
equation, because they too are compatible with the $n^i$. Indeed the
compatibility amounts to check that $n^i$ and $\tilde n^{(g)\,i}$ coincide 
modulo 2, which obviously remains true if signs are inserted. Moreover, the
identity occurs in the diagonal boundary conditions only, $\alpha =
\beta$, for which the signs cancel out.

On the contrary, in the case of a $Z_3$ symmetry, the insertion of signs
$\epsilon_\alpha$ does not yield sensible solutions (as far as the minimal models
are concerned). The reason is that some of the fields occur with multiplicity
two. Since the corresponding entries in $n^{(g)\,i}$ must be real combinations of
two third roots of unity, they can only be 2 or $-1$. Therefore, changing their
sign by inserting some $\epsilon_\alpha$ is not consistent.

Thus when the symmetry group is $Z_2$, there is a vast number of seemingly
acceptable solutions. These various solutions differ by the charges which are
assigned to the primary fields in mixed boundary conditions ($\alpha \neq
\beta$).  The freedom we have in choosing the $\epsilon_\alpha$ reflects the fact
that the charge normalization in mixed boundary conditions cannot be fixed a
priori, unlike what happens for diagonal boundary conditions, in which an 
identity occurs. 

One may try to derive more constraints on the charge normalizations by requiring
that the boundary charge assignments be compatible with {\it (i)} the charge
assignments in the bulk, and {\it (ii)} the boundary field operator product
coefficients. 

The first requirement is a condition on the way bulk fields close to a boundary
(taken to be $y=0$) expand in boundary fields \cite{cl,l}
\be
\phi_j(x+iy) \sim \sum_{{\rm b.c.}\;\alpha} \sum_k \; ^{(\alpha)}\!B_j^k \,
(2y)^{h_k-2h_j} \, \phi_k^{\alpha\alpha}(x),
\label{bbe}
\ee 
where the summation on $\alpha$ is over all boundary conditions, not just the
invariant ones. The $Z_2$ symmetry implies selection rules on the coefficients
since a bulk field of a given parity should expand in a combination of boundary
fields that transforms the same way. It means that the parity of the field
$\phi_k^{\alpha \alpha}$ must match that of $\phi_j$ for all invariant boundary
conditions $\alpha$ such that $^{(\alpha)}\!B_j^k \neq 0$.

Since these expansions involve fields in diagonal boundary conditions only, the
selection rules that follow are the same no matter what the signs
$\epsilon_\alpha$ are. This does not prove however that the selection rules are
indeed satisfied. For the diagonal models $(A,A)$, the coefficients 
$^{(\alpha)}\!B_j^k$ are known explicitly \cite{r}, and it would be interesting
to check directly that their values are consistent with the boundary charge
assignment found here.

The second check concerns the operator algebra of the boundary
fields themselves \cite{cl,l}
\be
\phi^{\alpha\beta}_i(x) \, \phi^{\beta\gamma}_j(x') \sim \sum_k \;
C^{(\alpha\beta\gamma)\,k}_{ij}(x-x')^{h_k-h_i-h_j}\,\phi^{\alpha\gamma}_k(x').
\label{bope}
\ee
Restricting to invariant boundary conditions $\alpha,\beta,\gamma$, the
discrete symmetry implies again selection rules which require that the charges
given by the matrices $n^{(g)\,i}$ provide a grading of the boundary 
fusion algebra\footnote{We leave aside the cases where some matrix elements
$n^{(g)\,i}_{\alpha,\beta}$ are zero without having the corresponding elements in
$n^i$ equal to zero. This happens when primary fields come in pairs of opposite
charge.}:
\be
C^{(\alpha\beta\gamma)\,k}_{ij} \neq 0 \quad \Longrightarrow \quad
n^{(g)\,i}_{\alpha,\beta} n^{(g)\,j}_{\beta,\gamma} = n^{(g)\,k}_{\alpha,\gamma}.
\label{bbf}
\ee
It is obvious that if the matrix coefficients $\tilde n^{(g)\,i}_{\alpha,\beta}$ 
satisfy  (\ref{bbf}), the same will be true of $\epsilon_\alpha \epsilon_\beta
\tilde n^{(g)\,i}_{\alpha,\beta}$, so that here too, these matrices are all 
consistent with the boundary operator product expansion (\ref{bope}), or else
none of them is. As the discrete symmetry is expected to occur, one can be
confident in the fact that the selection rules will be satisfied. We give below
examples of selection rules in the most explicit case, namely the diagonal models.
We have not shown in general that they are indeed satisfied, and as before, a proof
not based on symmetry arguments would be valuable.

In the diagonal models $(A,A)$, the boundary conditions are in one--to--one
correspondence with the chiral primary fields through their labelling
by two nodes $(a,b)$ taken in $A_{p-1}$ and $A_{q-1}$. As before, the 
boundary conditions $(a^*,b^*) = (p-a,q-b)$ and $(a,b)$ are to be identified.
Without loss of generality, one may thus assume that the first label
(the ``$r$--label'') is odd. 

The boundary operator product coefficients are known explicitly from \cite{r},
where they were proved to be equal to coefficients of the crossing matrices
(in a suitable normalization)
\be
C^{(\alpha\beta\gamma)\,k}_{ij} = F_{\beta,k}\left[\matrix{\alpha & \gamma \cr
i & j}\right].
\ee

Since for instance, an odd boundary field $\phi^{\alpha\alpha}_i$ cannot occur in
its fusion with itself, the corresponding crossing coefficient must
vanish. The verification that it does is non--trivial only when the chiral
field $i$ indeed occurs in its own bulk fusion (namely $N_{ii}^i \neq 0$), when
the primary field $i$ indeed occurs in the diagonal boundary conditions
($n^i_{\alpha,\alpha} \neq 0$ for $\alpha$ invariant under $Z_2$), and when it
is an odd field ($\tilde n^{(g)\,i}_{\alpha,\alpha} = -1$). All three conditions 
can be easily worked out, and yield
\be
F_{\alpha,i}\left[\matrix{\alpha & \alpha \cr
i & i}\right] = 0
\label{f}
\ee
for all $i=(r,s)$ such that $r,s$ are odd, $s=3 \bmod 4$, $r \leq (2p-1)/3$, $s
\leq (2q-1)/3$, and for all $\alpha=(a,q/2)$ such that $(r+1)/2 \leq a \leq p/2$.

The simplest example where such constraints arise is the tetracritical Ising model
$(A_4,A_5)$, in which (\ref{f}) implies (in terms of conformal weights)
\be
F_{{1 \over 15},{1 \over 15}}\left[{\scriptsize \matrix{{1 \over 15} & {1 \over
15} \cr {1 \over 15} & {1 \over 15}}}\right] = 
F_{{1 \over 15},{2 \over 3}}\left[{\scriptsize \matrix{{1 \over 15} & {1 \over 15}
\cr {2 \over 3} & {2 \over 3}}}\right] =
F_{{2 \over 3},{2 \over 3}}\left[{\scriptsize \matrix{{2 \over 3} & {2 \over 3}
\cr {2 \over 3} & {2 \over 3}}}\right] = 0.
\ee
More conditions can be derived in a generic diagonal model.

To summarize, the matrices $\tilde n^{(g)\,i}$ displayed in (\ref{nigf}) and
(\ref{deltas}) yield but a particular solution to Cardy's equation. For a $Z_3$
symmetry, they form the only consistent solution,
\be
n^{(g)\,i}_{\alpha,\beta} = \tilde n^{(g)\,i}_{\alpha,\beta}, \qquad (g^3=e),
\ee
whereas, in the case of a $Z_2$ symmetry, there are many more given by 
\be
n^{(g)\,i}_{\alpha,\beta} = \epsilon_\alpha \epsilon_\beta \tilde
n^{(g)\,i}_{\alpha,\beta}, \quad \epsilon_\alpha =
\pm 1 \qquad (g^2=e),
\label{nnn}
\ee
for arbitrary signs. The effect of these signs is to reverse (or to maintain,
depending to the value of $\epsilon_\alpha \epsilon_\beta$) the parity of all the
fields that occur in the sector of boundary conditions $\alpha,\beta$.

The ambiguity in the normalization of the $Z_2$ charges that arises due to these
signs must be resolved on physical grounds. As the interpretation of the
boundary fields is lacking in the general non--unitary model, it is not clear to
the author what the correct requirement should be. In this context, the specific
choice $\epsilon_\alpha=+1$ for all $\alpha$ is a minimal and natural one, as it
extends nicely the corresponding formula for $g=e$, and retains much of the graph
theoretic description. It also has the distinctive feature of producing matrices
$\tilde n^{(g)\,i}$ of constant sign, either totally positive or totally
negative\footnote{There is another solution in terms of matrices of constant
sign, which is obtained by substituting $-G^\sigma$ for $G^\sigma$ in the formula
(\ref{nigf}) giving $\tilde n^{(g)\,i}$. The substitution has no effect when
$G^\sigma=A_1$, since the associated adjacency matrix is the number zero, while
in the other cases, it causes the matrices $\tilde n^{(g)\,i}$ to be multiplied by
$(-1)^{s+1}$. This sign can be seen to be in the line of the previous discussion,
because it is equal to $(-1)^{s+1} = \epsilon_\alpha \epsilon_\beta$ with
$\epsilon_\alpha = (-1)^{b+1}$ if $\alpha=(a,b)$. The existence of this solution 
is a consequence of a non--trivial automorphism of the graph $G^\sigma$.}. However
in view of what follows, this may not be the correct choice. 

In a unitary model, the ground state of every sector is expected to
be invariant under the symmetry group, on account of the Perron--Frobenius
theorem applied to the transfer matrix. This provides a well--defined criterion
to fix the normalization of the charges, and therefore the physical value of
the signs $\epsilon_\alpha$. We will use this criterion as a guide, in order to see
if a particular set of values $\epsilon_\alpha$ emerges from this point of view.

%%%%%%%%%%%%%%%%%%%%%%%%%%%%%%%%%%%%%%%%%%%%%%%%%%%%%%%%%%%%%%%%%%%%%%%%%%%%%%

\section{Unitary models}

In this last section, we explore the possibility of fixing the value of the signs
$\epsilon_\alpha$ by using the criterion we have just mentioned: if the continuum
limit is smooth enough, it is expected that the consequences of the
Perron--Frobenius theorem on the finite--dimensional transfer matrix be maintained
in the corresponding conformal field theory. In particular, for all invariant
boundary conditions, the ground state of the Hamiltonian $H_{\alpha,\beta}$
(the primary field of lowest conformal dimension in ${\cal
H}_{\alpha,\beta}$) should be non--degenerate and (hence) invariant under the
symmetry group. In short, we will call this the Perron--Frobenius (PF)
criterion.  As already said, it is automatically satisfied in the diagonal boundary
conditions.

Thus we look for a set of $\epsilon_\alpha$ such that the $Z_2$ charge assignment
meet the PF criterion. Incidentally, when the symmetry group is
$Z_3$, there is only one consistent charge assignment (see the previous section).
In that case, we will merely check whether the PF criterion is satisfied. 

The outcome of this investigation is somewhat surprising. The unitary diagonal
models are the only ones where the PF criterion can be met, for a unique choice of
the $\epsilon_\alpha$'s. In all other unitary models, there is no way in which
it can be fulfilled, if one insists that it be valid in all sectors.
A physical interpretation of this will be proposed\footnote{I am indebted to
Gerard Watts for a clarifying discussion about this issue.}. Nonetheless, for
all those models but two, we will see that a unique set of $\epsilon_\alpha$'s is
singled out by demanding a minimal violation of the PF criterion.

We recall that the conformal weight of a primary field labelled by $i=(r,s)$ is
equal to
\be
h_{r,s} = {(qr-ps)^2-(p-q)^2 \over 4pq}.
\ee
Throughout this section, we will take $p$ odd and $q=p \pm 1$ even. Then the
smallest conformal weights correspond, in ascending order, to $i=(1,1), (2,2),
(3,3), \ldots$.

\subsection{The unitary series (A,A)}

The only boundary primary fields that occur in the diagonal models have their
$s$--label odd (see (\ref{aa})). Since the identity (1,1) does not appear in mixed
boundary conditions, the primary with the lowest weight that can possibly occur in
mixed boundary conditions corresponds to (3,3), and consequently, the off--diagonal
entries of 
\be
n^{(g) \, (3,3)}_{\alpha,\beta} = n^{(g) \, (3,3)}_{(a_1,{q \over 2}),(a_2,{q
\over 2})} = -\epsilon_{a_1}\,\epsilon_{a_2}\,U_2(T_{p-1 \over 2})_{a_1,a_2}
\ee
must be positive. The off--diagonal matrix coefficients $U_2(T)_{a_1,a_2}$ equal 1
if $|a_1-a_2|=2$ or if $\{a_1,a_2\}=\{(p-3)/2,(p-1)/2\}$, and 0 otherwise (it
counts the number of paths of length 2 going from $a_1$ to $a_2$ on the graph
$T_{(p-1)/2}$). Thus one obtains the condition $\epsilon_{a_1}
\epsilon_{a_2}=-1$ for all these pairs. This fixes the vector $\epsilon_a$ in a
unique way (up to a global sign that does not matter) as
\be
\epsilon_a = (\ldots,+1,+1,-1,-1,+1,+1,-1,-1,+1).
\label{epa}
\ee

For these specific signs, one may then verify that in the remaining mixed
boundary sectors (those for which (3,3) does not occur), the field of lowest
weight has indeed a parity +1 (zero charge). To do that, one can first observe
that any mixed boundary sector has its field of lowest weight in $\{(r,s) \;:\;
3 \leq r=s \;{\rm odd} \leq p-2\}$. Next point is to note that
$U_{r-1}(T)_{a_1,a_2}$ is zero unless the nodes $a_1,a_2$ can be related by a
path of length $r-1$. If the two nodes cannot be connected by a shorter path, it
follows from (\ref{epa}) that $\epsilon_{a_1} \epsilon_{a_2} = (-1)^{(r-1)/2}$, 
so that the numbers
\be
n^{(g) \, (r,r)}_{(a_1,{q \over 2}),(a_2,{q \over 2})} = 
\epsilon_{a_1} \epsilon_{a_2} (-1)^{(r-1)/2} U_{r-1}(T)_{a_1,a_2}
\ee
are positive (or zero). That $a_1$ and $a_2$ can be connected by a shorter path
means that the field $(r,r)$ is not the primary with the lowest weight in that
sector, and we are back to the first case.

Since the PF criterion can be satisfied in all sectors for a unique set of
$\epsilon_\alpha$'s, it is tempting to conjecture that these are the correct
physical values. The charge content in the various sectors of the unitary diagonal
models would then be given by 
\be
n^{(g) \, i}_{(a_1,{q \over 2}),(a_2,{q \over 2})} = \epsilon_{a_1} \,
\epsilon_{a_2} \, \tilde n^{(g) \, i}_{(a_1,{q \over 2}),(a_2,{q \over 2})},
\ee
with the signs (\ref{epa}), and the $\tilde n^{(g)\,i}$ as in (\ref{aa}).

\subsection{The unitary series (A,D)}

The same calculations can be carried out for the unitary models of the $(A,D)$
series, with however different results. To illustrate it most clearly, we will
start with the simplest case, namely $(A_4,D_4)$, corresponding to the
critical 3--Potts model ($p=5$, $q=6$). 

A set of $Z_2$--symmetric boundary conditions is provided\footnote{The model has 
eight conformally invariant boundary conditions which are invariant under a $Z_2$,
but not under the same $Z_2$. One finds three groups of four boundary conditions
that are simultaneously invariant under the same $Z_2$. They clearly correspond
to the three conjugate $Z_2$ subgroups of $S_3$, the automorphism group of
$D_4$.} by the so--called A, BC, Free and New \cite{aos}. They correspond
respectively to the nodes (1,1), (2,1), (1,2) and (2,2). (Free and New, being
fully invariant under $S_3$, must correspond to $b=2$, which is the only node of
$D_4$ invariant under $S_3$.) Together they define 10 different sectors.

It is not difficult to find the field with lowest weight in each of these
sectors, and then compute the parity they are assigned by the matrices
$\tilde n^{(g)\,i}$ computed in Section \ref{sec:expl}. Writing these in
a matrix $\tilde M$, one obtains (indices are A, BC, Free, New)
\bea
\tilde M_{\alpha,\beta} &=& \Big(\tilde n^{(g)\,imin}_{\alpha,\beta} \;:\; \min_{i
\in {\cal H}_{\alpha,\beta}} h_i = h_{imin}\Big)_{\alpha,\beta} \nonumber\\
&=& {\footnotesize \pmatrix{+1 & -1 & +1 & -1 \cr -1 & +1 & -1 & -1 \cr 
+1 & -1 & +1 & 0 \cr -1 & -1 & 0 & +1}}.
\eea
The zeros are due to the partition function (superscripts are the conformal
weights)
\be
Z_{\rm Free,New} = 2\,\chi_{3,3}^{1/15} + \chi_{3,5}^{2/5} + \chi_{3,1}^{7/5},
\ee
which shows that the ground state in that sector is doubly degenerate, the two
states having opposite parities. 

The above matrix makes clear that the charge assignment implied by $\tilde
n^{(g)\,i}$ does not satisfy the PF criterion in all sectors, either because the
ground state is not invariant, or because it is degenerate. One may try to
find values for $\epsilon_\alpha$ that render the non--degenerate ground states
invariant, but one easily sees that it is not possible: no values for
$\epsilon_\alpha$ can be found so that $\tilde M_{\alpha,\beta} \geq 0$ for all
$\alpha,\beta$.

One can relax our demands by looking for a set of $\epsilon_\alpha$ which minimizes
the number of sectors that violate the PF criterion. One then finds that the
minimal number of such sectors, which we call non--PF, is equal to 
\be
N_{\rm non-PF} = 2.
\ee
This number is realized for $\epsilon_\alpha =  (+1,-1,+1,-1) = (+1,-1)_a \otimes
(1,1)_b$, the two non--PF sectors being BC,New and Free,New. Indeed for these
$\epsilon_\alpha$, one obtains 
\be
\epsilon_\alpha \epsilon_\beta \tilde M_{\alpha,\beta} = 
{\footnotesize \pmatrix{+1 & +1 & +1 & +1 \cr +1 & +1 & +1 & -1 \cr 
+1 & +1 & +1 & 0 \cr +1 & -1 & 0 & +1}}.
\ee
Let us also notice that if one excludes just one boundary condition, namely
``New'', the expected consequences of the PF theorem are indeed verified. Thus in
this case, the minimal number of boundary conditions that have to be excluded for
this to be true is equal to 1.

Finally one may note that $\epsilon_\alpha = (+1,-1,+1,+1)$ share the same
properties, the two non--PF sectors being now A,New and Free,New. 

In any case, one must conclude that the transfer matrix, in certain sectors of
boundary conditions, does not satisfy the conditions of the PF theorem. There can
be only two reasons for that: either the transfer matrix is not
irreducible\footnote{The unicity of the largest eigenvalue is guaranteed only for
non--negative primitive matrices \cite{ms}. Under mild assumptions on the transfer
matrix, its irreducibility is sufficient.}, or else it contains negative entries,
implying that some of the boundary Boltzmann weights are negative (or both).

That the first condition fails is unlikely because the periodic transfer matrix is
irreducible and because the boundary conditions are undecomposable. So one should
favour the second alternative, which points to the unphysical nature of some of the
boundary conditions, their classical description requiring negative Boltzmann
weights. We note that a boundary condition $\alpha$ which is described by negative
Boltzmann weights does not necessarily lead to unphysical (negative, non--PF)
partition functions. Whether or not this is the case depends on which other 
boundary condition is associated with $\alpha$.

The appearance of negative classical boundary Boltzmann weights to describe the
New boundary condition in the critical 3--Potts model has been discussed in
\cite{aos}, and is confirmed by the explicit calculation of the critical boundary
weights \cite{p}.

As we shall see, what is true in the 3--Potts model is true in all unitary
models of the $(A,D)$ series. No values for the $\epsilon_\alpha$'s exist which
make all sectors to satisfy the PF criterion, but a suitable choice, unique, 
contrary to the above case, of $\epsilon_\alpha$ minimizes the number of sectors
which do not satisfy it. As above we will take the point of view that these
features are the signal that a certain number of boundary conditions are
unphysical, because they require negative Boltzmann weights for their classical
description. 

We have not carried out the analysis of the whole series, but instead we have
investigated the first eight models, up to $p=13$ and $q=12$, with the following
results.

In each of these models, we have determined the minimal number $N_{\rm unphys}$ of
boundary conditions that must be disregarded in order to satisfy the PF criterion
in all the sectors involving the remaining ones. This uniquely singles out a set of
boundary conditions, which naturally qualifies as the set of unphysical boundary
conditions. This also determines unique values of the $\epsilon_\alpha$ for the
physical ones. The values of $\epsilon_\alpha$ for the unphysical $\alpha$ are
then fixed (uniquely, except in the 3--Potts model) by requiring a minimal number
of non--PF sectors (which necessarily correspond to one or two unphysical boundary
conditions). That minimal number is denoted by $N_{\rm non-PF}$. The results are as
follows. 

In the model $(A_{p-1},D_{q/2+1})$ (we have looked at the eight models
corresponding to $6 \leq q \leq 12$), the number $N_{\rm unphys}(p,q)$ only depends
on the rank of the $D$ factor. It increases rather quickly since it is equal to 1,
3, 6 and 10 for the two models involving the algebra $D_4$, $D_5$, $D_6$ and $D_7$
respectively. We found that the unphysical boundary conditions form the set (the
labelling of the nodes is as in the figure of Section \ref{sec:sym})
\be
\{\alpha = (a,b) \in T_{p-1 \over 2} \times A_{{q \over 2}-1} \;:\; a+b \geq
{\textstyle {p+3 \over 2}}\}.
\label{excl}
\ee
Moreover, the signs which make the number of non--PF sectors minimal are unique and 
given by
\bea
\epsilon_\alpha &=& (+1,-1,+1,-1,\ldots)_a \otimes (1,1,1,\ldots)_b \nonumber\\
&=& (-1)^{a+1}, \qquad \alpha=(a,b).
\label{epd}
\eea
As pointed out above, in the model $(A_4,D_4)$, there is another solution
$\epsilon_\alpha = (+1,-1,+1,+1)$, which however appears to contradict the duality
relations (see below). 

We have determined $N_{\rm non-PF}$ by mere counting, and found that it 
equals 2, 3, 11, 15, 36, 46, 89, 109 for the first eight models, ordered as
$(A_4,D_4)$, $(A_6,D_4)$, $(A_6,D_5)$, ... (By symmetry, the sectors
$(\alpha,\beta)$ and $(\beta,\alpha)$ are identical and count for one.) 

These results strongly suggest the general pattern in which the number of boundary
conditions in (\ref{excl}) equals a binomial coefficient
\be
N_{\rm unphys}(A,D_{{q \over 2}+1}) = {{q \over 2}-1 \choose 2}.
\ee
This is a large number since essentially half the invariant boundary conditions
would have to be discarded as classically unphysical. A bit more of numerology
also shows that the number of non--PF sectors fits the simple formula 
\bea
N_{\rm non-PF}(A_{q\mp 1-1},D_{{q \over 2}+1}) &=& \nonumber\\
&& \hspace{-1.5cm} \big\{({\textstyle {q-2 \over 4}})^4\big\} + 
{\textstyle {q(q-2)(q \mp 2)(q-4) \over 192}},
\label{unphys}
\eea
where $\{x\}$ is the integer closest to $x$. The two numbers in the r.h.s. of the
previous equation have separately a well--defined meaning: the first one is the
number of sectors where the ground state is non--degenerate but odd under the
$Z_2$ symmetry, while the second one gives the number of sectors where the ground
state is doubly degenerate.  

The reader may wish to check the above assertions in a less simple instance than
the Potts model. A good example is to consider the $(A_6,D_5)$ model, for which one
computes (in the tensor product basis)
\be
\tilde M_{\alpha,\beta} = 
{\footnotesize \pmatrix{
+1 & -1 & +1 & +1 & -1 & +1 & +1 & -1 & +1  \cr  
-1 & +1 & -1 & -1 & +1 & -1 & -1 & +1 & +1 \cr 
+1 & -1 & +1 & +1 & -1 & -1 & +1 & +1 & +1 \cr 
+1 & -1 & +1 & +1 & -1 & +1 & +1 & -1 & 0 \cr  
-1 & +1 & -1 & -1 & +1 & +1 & -1 & 0 & -1 \cr 
+1 & -1 & -1 & +1 & +1 & +1 & 0 & -1 & -1 \cr 
+1 & -1 & +1 & +1 & -1 & 0 & +1 & 0 & 0 \cr  
-1 & +1 & +1 & -1 & 0 & -1 & 0 & +1 & 0 \cr 
+1 & +1 & +1 & 0 & -1 & -1 & 0 & 0 & +1}}. 
%   
%   +1 & -1 & +1 & -1 & +1 & -1 & +1 & -1 & +1 & -1 & +1 & -1 \cr  
%   -1 & +1 & -1 & +1 & -1 & +1 & -1 & +1 & -1 & +1 & -1 & +1 \cr 
%   +1 & -1 & +1 & -1 & +1 & -1 & +1 & -1 & +1 & -1 & +1 & +1 \cr  
%   -1 & +1 & -1 & +1 & -1 & +1 & -1 & -1 & -1 & +1 & +1 & +1 \cr 
%   +1 & -1 & +1 & -1 & +1 & -1 & +1 & -1 & +1 & -1 & +1 & 0 \cr  
%   -1 & +1 & -1 & +1 & -1 & +1 & -1 & +1 & -1 & +1 & -1 & 0 \cr 
%   +1 & -1 & +1 & -1 & +1 & -1 & +1 & +1 & +1 & -1 & 0 & -1 \cr  
%   -1 & +1 & -1 & -1 & -1 & +1 & +1 & +1 & 0 & 0 & -1 & -1 \cr 
%   +1 & -1 & +1 & -1 & +1 & -1 & +1 & 0 & +1 & -1 & 0 & 0 \cr  
%   -1 & +1 & -1 & +1 & -1 & +1 & -1 & 0 & -1 & +1 & 0 & 0 \cr 
%   +1 & -1 & +1 & +1 & +1 & -1 & 0 & -1 & 0 & 0 & +1 & 0 \cr  
%   -1 & +1 & +1 & +1 & 0 & 0 & -1 & -1 & 0 & 0 & 0 & +1 }}. \nonumber\\
%
\ee

The values of $\epsilon_\alpha$ mentioned in (\ref{epd}) are nothing but the first
line of $\tilde M_{\alpha,\beta}$, and the boundary conditions to discard label
the rows and columns 6, 8 and 9, which correspond, in terms of the fixed point
graph $T_3 \times  A_3$, to the pairs of nodes $(a,b) = (3,2),(2,3)$ and $(3,3)$,
as announced in (\ref{excl}). There are 6 zeros in the upper triangular part of 
$\tilde M_{\alpha,\beta}$, which is the value of the second summand of
(\ref{unphys}).

All this leads to the reasonable guess that (\ref{epd}) might give the
correct physical values of the $\epsilon_\alpha$'s. Inserted in (\ref{nnn}), it not
only determines the parities of all primaries in the sectors where the PF criterion
is satisfied, but it also points to the boundary conditions that can have a
problematic lattice interpretation. These conjectural statements must be confirmed
or dismissed by the explicit calculation of the boundary Boltzmann weights. The
results obtained so far seem to give some support to our conjecture \cite{betal}.

Assuming this conjecture, it is not difficult to give an explicit formula for the
parities. From (\ref{nigf}), (\ref{deltas}) and (\ref{epd}), they are determined
from 
\bea
n^{(g)\,i}_{(a_1,b_1),(a_2,b_2)} &=& (-1)^{a_1+a_2+r+1} \times \nonumber\\
&& \hspace{-7mm} U_{r-1}(T_{(p-1) \over 2})_{a_1,a_2} \, U_{s-1}(A_{{q \over
2}-1})_{b_1,b_2}.
\eea
The matrices $U_{r-1}(T_{(p-1)/2})$ are all positive,  unlike
the $U_{s-1}(A_{q/2-1})$, which are positive for $s<q/2$,  negative for
$s>q/2$, and identically zero for $s=q/2$, on account of $U_{q-s-1}(A_{q/2-1}) =
- U_{s-1}(A_{q/2-1})$. 

Putting all these observations together, one can conclude that the paired fields
have opposite $Z_2$ parities within each pair (as already pointed out), and that the
parity of an unpaired field in the sector of boundary conditions $\alpha,\beta$ is
equal to
\be
g(\phi_i^{\alpha\beta}) = \cases{
(-1)^{a_1+a_2+r+1} \, \phi_i^{\alpha\beta} & if $s<q/2$, \cr
\noalign{\smallskip}
(-1)^{a_1+a_2+r} \, \phi_i^{\alpha\beta} & if $s>q/2$.}
\ee

In the critical 3--Potts model for instance, one finds the following frustrated
partition functions (in terms of the conformal weights)
\bea
&& Z^g_{\rm A,A} = \chi_{0} - \chi_{3}, \\
&& Z^g_{\rm A,BC} = \chi_{2/5} - \chi_{7/5}, \\
&& Z^g_{\rm A,Free} = \chi_{1/8} - \chi_{13/8}, \\
&& Z^g_{\rm BC,BC} = \chi_{0} - \chi_{3} - \chi_{2/5} + \chi_{7/5}, \\
&& Z^g_{\rm BC,Free} = \chi_{1/40} - \chi_{21/40},\\
&& Z^g_{\rm Free,Free} = \chi_{0} - \chi_{3} + \chi_{2/3} - \chi_{2/3^+},\\
&& Z^g_{\rm A,New} = \chi_{1/40} - \chi_{21/40}, \label{anew} \\
&& Z^g_{\rm New,New} = \chi_{0} - \chi_{3} - \chi_{2/5} + \chi_{7/5} \nonumber\\
&& \hspace{1.5cm} + \chi_{2/3} - \chi_{2/3^+} + \chi_{1/15} - \chi_{1/15^+}.
\eea

These functions are computed using the $\epsilon_\alpha$'s given in (\ref{epd}),
and appear to be consistent with the duality of the model \cite{aos}. For
instance, the equality 
\be
Z_{\rm BC,Free} = Z_{\rm A,New}
\label{dual}
\ee 
is maintained for the frustrated partition functions, while 
\be
Z_{\rm Free,Free} = Z_{\rm A,A} + Z_{\rm A,B} + Z_{\rm A,C} 
\ee
becomes $Z^g_{\rm Free,Free} = Z^g_{\rm A,A}$ since $Z^g_{\rm A,B} = Z^g_{\rm A,C}
= 0$. 

The use of the other solution $\epsilon_\alpha = (+1,-1,+1,+1)$ has the effect of
multiplying by $-1$ the partition functions of all sectors with one ``New'', 
so that $Z^g_{\rm A,New}$ would be minus the expression in (\ref{anew}),
contradicting the duality relation (\ref{dual}). 
 
There is a $Z_3$ symmetry in two models only, namely the critical and tricritical
3--Potts models $(A_4,D_4)$ and $(A_6,D_4)$. They possess respectively 2 (``Free''
and ``New'') and 3 invariant boundary conditions, namely $\alpha=(a,2)$ for $a$ a
node of
$T_2$ and $T_3$. The relevant $\tilde M$ matrices are equal to
\be
\tilde M_{\alpha,\beta} = \footnotesize{\pmatrix{+1 & -1 \cr -1 & +1}} 
\quad {\rm and} \quad 
\footnotesize{\pmatrix{+1 & +1 & -1 \cr +1 & +1 & -1 \cr -1 & -1 & +1}},
\ee
where a $-1$ sign indicates that the corresponding sector has two degenerate ground
states, of opposite and non--zero charge (none of them is invariant under the
$Z_3$).

In the first case (the $(A_4,D_4)$ model), it is the second boundary condition
$(2,2)$ (i.e. ``New'') that appears to be unphysical, while in the second case, it
is the third boundary condition $(3,2)$. This should not be surprising since they
are precisely the boundary conditions which were unphysical from the $Z_2$ point
of view: from (\ref{excl}), $\alpha=(a,2)$ was to be discarded if $a+2 \geq
(p+3)/2$, that is, if $a=(p-1)/2$. Therefore, the boundary conditions which were
causing problems for the $Z_2$ charges also cause problems for the $Z_3$ charges. 

\subsection{The unitary models (A,E$_{\bf 6}$)}

We will content ourselves with making a few comments on the two unitary models 
$(A_{10},E_6)$ and $(A_{12},E_6)$ ($p=11$ or 13, and $q=12$). 

As we have said above, the models involving the $E_6$ algebra have the peculiarity
of possessing primary fields that occur with multiplicity 1, 2 and 3. It turns out
that the same is true of the ground state in various sectors. Let us examine in
some detail the simplest model $(A_{10},E_6)$.

That model possesses 10 invariant boundary conditions, labelled as $\alpha=(a,b)$
with $a=1,2,\ldots,5$ a node of $T_5$, and $b=3,6$ a node of the $A_2$ subgraph of
$E_6$, fixed by its non--trivial automorphism. One can compute as before the matrix
$\tilde M_{\alpha,\beta}$ which collects those entries of
$\tilde n^{(g)\,i}_{\alpha,\beta}$ for which $i$ is the lowest weight primary in
the sector $\alpha,\beta$. The result is
\bea
&& \tilde M_{\alpha,\beta} = \nonumber\\
&& {\footnotesize \pmatrix{
+1 & 0 & 0 & +1^* & -1^* & +1 & -1 & -1 & +1 & 0 \cr
0 & +1 & +1^* & 0 & -1^* & -1 & -1 & -1 & 0 & +1 \cr
0 & +1^* & +1 & -1^* & 0 & -1 & -1 & 0 & -1 & +1 \cr
+1^* & 0 & -1^* & +1 & 0 & +1 & 0 & -1 & +1 & -1 \cr
-1^* & -1^* & 0 & 0 & +1 & 0 & +1 & +1 & -1 & -1 \cr 
+1 & -1 & -1 & +1 & 0 & +1 & -1 & -1 & +1 & -1 \cr
-1 & -1 & -1 & 0 & +1 & -1 & +1 & +1 & -1 & -1 \cr
-1 & -1 & 0 & -1 & +1 & -1 & +1 & +1 & -1 & -1 \cr 
+1 & 0 & -1 & +1 & -1 & +1 & -1 & -1 & +1 & -1 \cr
0 & +1 & +1 & -1 & -1 & -1 & -1 & -1 & -1 & +1}}, \nonumber\\
\eea
where the stars mean that the ground state in the corresponding sector is three
times degenerate, the number $\pm 1$ being the sum of the three parities. As
before, a zero indicates that there are two degenerate ground states with opposite
parity. 

We can repeat what we did for the $(A,D)$ series, and look for a set of 
$\epsilon_\alpha$ which minimizes the violation of the PF criterion.  

By varying the $\epsilon_\alpha$, one finds that the minimal number of
non--PF sectors is equal to 21, and that the non--PF sectors have at least one 
boundary condition in the set
\be
\{(2,3), (3,3), (4,3), (5,3), (5,6)\}
\ee
in terms of the nodes of $T_5 \times A_2$ (they correspond to the rows and columns
2, 3, 4, 5, 10). So these five boundary conditions can presumably be called
unphysical in the sense of the previous subsection. Hence
\be
N_{\rm unphys}(A_{10},E_6) = 5, \quad N_{\rm non-PF}(A_{10},E_6) = 21.
\ee

There are four solutions for the $\epsilon_\alpha$'s for which these values can be
realized. Among them, the most symmetrical one is $\epsilon_\alpha =
(+1,-1,-1,+1,-1) \otimes (1,1)$. 

The other model $(A_{12},E_6)$ is similar. One finds 
\be
N_{\rm unphys}(A_{12},E_6) = 5, \quad N_{\rm non-PF}(A_{12},E_6) = 27.
\ee
The presumably unphysical boundary conditions correspond to the nodes (3,3), (4,3),
(5,3), (6,3), (6,6) of $T_6 \times A_2$. The signs for which these numbers are
reached are unique and given by $\epsilon_\alpha = (+1,-1,+1,+1,-1,+1) \otimes
(1,1)$.

\acknowledgments
It is a pleasure to thank Jean-Bernard Zuber for an encouraging and stimulating
exchange, and for reading the manuscript. I also wish to thank Paul Pearce who has
informed me of the preliminary results that he and his collaborators have obtained
on boundary Boltzmann weights, Ingo Runkel for pointing out misprints in the draft,
and Gerard Watts for a very fruitful discussion.

%%%%%%%%%%%%%%%%%%%%%%%%%%%%%%%%%%%%%%%%%%%%%%%%%%%%%%%%%%%%%%%%%%%%%%%%%%%%%%

\end{document}